\documentclass[aip,reprint]{revtex4-1}

\usepackage{graphicx}
\usepackage{dcolumn}
\usepackage{bm}
\usepackage{xcolor}

\usepackage[utf8]{inputenc}
\usepackage[T1]{fontenc}
\usepackage{mathptmx}
\usepackage{etoolbox}

\draft 

\begin{document}


\title[First observation of quantum oscillations by transport measurements in semi-destructive pulsed magnetic fields up to 125 T]{First observation of quantum oscillations by transport measurements in semi-destructive pulsed magnetic fields up to 125 T} 



\author{Maxime Massoudzadegan}
\altaffiliation{equally contributed}
\affiliation{ LNCMI-EMFL, CNRS UPR3228, Univ. Grenoble Alpes,
Univ. Toulouse, INSA-T, Toulouse, France}
\affiliation{Institut Quantique, Département de physique,
Université de Sherbrooke, Sherbrooke, Québec, Canada}
\affiliation{Université de Sherbrooke – CNRS / IRL Frontières Quantiques, Sherbrooke, Canada}

\author{Sven Badoux}
\altaffiliation{equally contributed}
\affiliation{ LNCMI-EMFL, CNRS UPR3228, Univ. Grenoble Alpes,
Univ. Toulouse, INSA-T, Toulouse, France}

\author{Nicolas Bruyant}
\affiliation{ LNCMI-EMFL, CNRS UPR3228, Univ. Grenoble Alpes,
Univ. Toulouse, INSA-T, Toulouse, France}

\author{Ildar Gilmutdinov}
\affiliation{ LNCMI-EMFL, CNRS UPR3228, Univ. Grenoble Alpes,
Univ. Toulouse, INSA-T, Toulouse, France}

\author{Isaac Haik-Dunn}
\affiliation{ LNCMI-EMFL, CNRS UPR3228, Univ. Grenoble Alpes,
Univ. Toulouse, INSA-T, Toulouse, France}

\author{Gabriel de Oliveira Rodrigues}
\affiliation{ LNCMI-EMFL, CNRS UPR3228, Univ. Grenoble Alpes,
Univ. Toulouse, INSA-T, Toulouse, France}

\author{Nuno Lourenco Prata}
\affiliation{ LNCMI-EMFL, CNRS UPR3228, Univ. Grenoble Alpes,
Univ. Toulouse, INSA-T, Toulouse, France}

\author{Abdelaziz Zitouni}
\affiliation{ LNCMI-EMFL, CNRS UPR3228, Univ. Grenoble Alpes,
Univ. Toulouse, INSA-T, Toulouse, France}

\author{Marc Nardone}
\affiliation{ LNCMI-EMFL, CNRS UPR3228, Univ. Grenoble Alpes,
Univ. Toulouse, INSA-T, Toulouse, France}

\author{Oleksiy Drashenko}
\affiliation{ LNCMI-EMFL, CNRS UPR3228, Univ. Grenoble Alpes,
Univ. Toulouse, INSA-T, Toulouse, France}

\author{Oliver Portugall}
\affiliation{ LNCMI-EMFL, CNRS UPR3228, Univ. Grenoble Alpes,
Univ. Toulouse, INSA-T, Toulouse, France}

\author{Steffen Wiedmann}
\affiliation{High Field Magnet Laboratory (HFML-EMFL) and Institute for Molecules and Materials, Radboud University, Toernooiveld 7, 6525 ED Nijmegen, The Netherlands}

\author{Benoît Fauqué}
\affiliation{JEIP, USR 3573 CNRS, Collège de France, PSL Research University,
, Paris, France}

\author{David Vignolles}
\email[]{david.vignolles@lncmi.cnrs.fr}
\affiliation{ LNCMI-EMFL, CNRS UPR3228, Univ. Grenoble Alpes,
Univ. Toulouse, INSA-T, Toulouse, France}
\affiliation{Université de Sherbrooke – CNRS / IRL Frontières Quantiques, Sherbrooke, Canada}

\author{Bertrand Reulet}
\email[]{bertrand.reulet@USherbrooke.ca}
\affiliation{Institut Quantique, Département de physique,
Université de Sherbrooke, Sherbrooke, Québec, Canada}
\affiliation{Université de Sherbrooke – CNRS / IRL Frontières Quantiques, Sherbrooke, Canada}

\author{Cyril Proust}
\email[]{cyril.proust@lncmi.cnrs.fr}
\affiliation{ LNCMI-EMFL, CNRS UPR3228, Univ. Grenoble Alpes,
Univ. Toulouse, INSA-T, Toulouse, France}
\affiliation{Université de Sherbrooke – CNRS / IRL Frontières Quantiques, Sherbrooke, Canada}


\date{\today}

\begin{abstract}
High magnetic fields have proven instrumental in exploring the physical properties of condensed matter, leading to groundbreaking discoveries such as the quantum Hall effect in 2D heterostructures and quantum oscillations in cuprate superconductors. 
The ability to conduct precise measurements at progressively higher magnetic fields continues to push the frontiers of knowledge and enable new discoveries. 
In this work, we present the development of a microwave technique for performing two-point transport measurements in semi-destructive pulsed magnetic fields (up to 125~T) and at low temperatures (down to 1.5~K) with unprecedented sensitivity. 
This new setup was tested on a variety of samples. We present results on the metal-insulator transition in InAs and we report notably the first observation of Shubnikov-de-Haas oscillations in WTe$_{2}$ at magnetic fields beyond 100~T.
\end{abstract}


\maketitle 

\section{Introduction}
High magnetic fields have been widely used in condensed matter physics to probe the electronic properties of materials. In the 50s and the 60s, high magnetic fields has enabled the study of fermiology of metals through the observation of quantum oscillations\cite{Shoenberg}. In case of superconductors, it allows to quench superconductivity in order to study the normal state \cite{Boebinger1996,Cooper2009}. Thanks to the improvement of signal-to-noise ratio in pulsed field measurements, the first observation of quantum oscillations in YBa$_2$Cu$_3$O$_y$ has given a new twist to the understanding of the normal state of hole-doped cuprate superconductors \cite{Doiron2007}. High magnetic fields help also to reveal new phases of matter, such as the fractional quantum Hall effect \cite{Stormer1999} and the field-induced superconductivity in U-based heavy fermions superconductors\cite{Aoki2013}. Last but not least, high magnetic fields enable to reach the quantum limit in diluted metals -- all the electrons are in their lowest Landau level -- and can trigger phase transition, such as in graphite \cite{Fauque2013}. The highest fields achievable today are produced by pulsed magnets and reach approximately 100 T\cite{Nguyen2016,Beard2018}. Pushing the limit of higher magnetic fields is crucial to allow the study of new materials and discover new phenomena. \\
To exceed this limit, the only viable approach currently involves a semi-destructive setup, in which a single-turn coil is destroyed during the pulse, while the cryostat and the probe containing the sample remain intact.
At the same time, this requires the development of experimental setup with a sufficiently high signal-to-noise ratio to observe the desired effect. 
Previous setups to measure the electrical resistivity of metals in single turn coil installation have been developed, including four-point measurements \cite{Miura2002}, millimeter-wave transmission \cite{Shimamoto1998}, radio frequency (RF) transmission \cite{Sekitani2007} or reflection \cite{Nakamura2018,Shitaokoshi2023}. However, up to now, the signal-to-noise ratio did not allow the observation of quantum oscillations in very strong magnetic field and this was the target of our experimental development.\\
In this article, we report the development of microwave transport technique in strong magnetic fields up to 125~T at cryogenic temperatures down to 1.5~K using the single-turn coil installation at the LNCMI-Toulouse. We reached an excellent signal-to-noise ratio that allow to observe the metal-insulator transition in InAs when the quantum limit is reached \cite{Jaoui2020}.This transition is highly sensitive to the temperature and the comparison with non-destructive pulsed field data was used to estimate the heating effects in the semi-destructive installation. Building on this initial success, we report the first observation of quantum oscillation up to 125~T in the semi-metal WTe$_{2}$, in excellent agreement with previously published data at lower fields \cite{Linnartz2022}.



\section{Methods}
\subsection{Megagauss installation and constraints}

\begin{figure}
\includegraphics[scale=0.32]{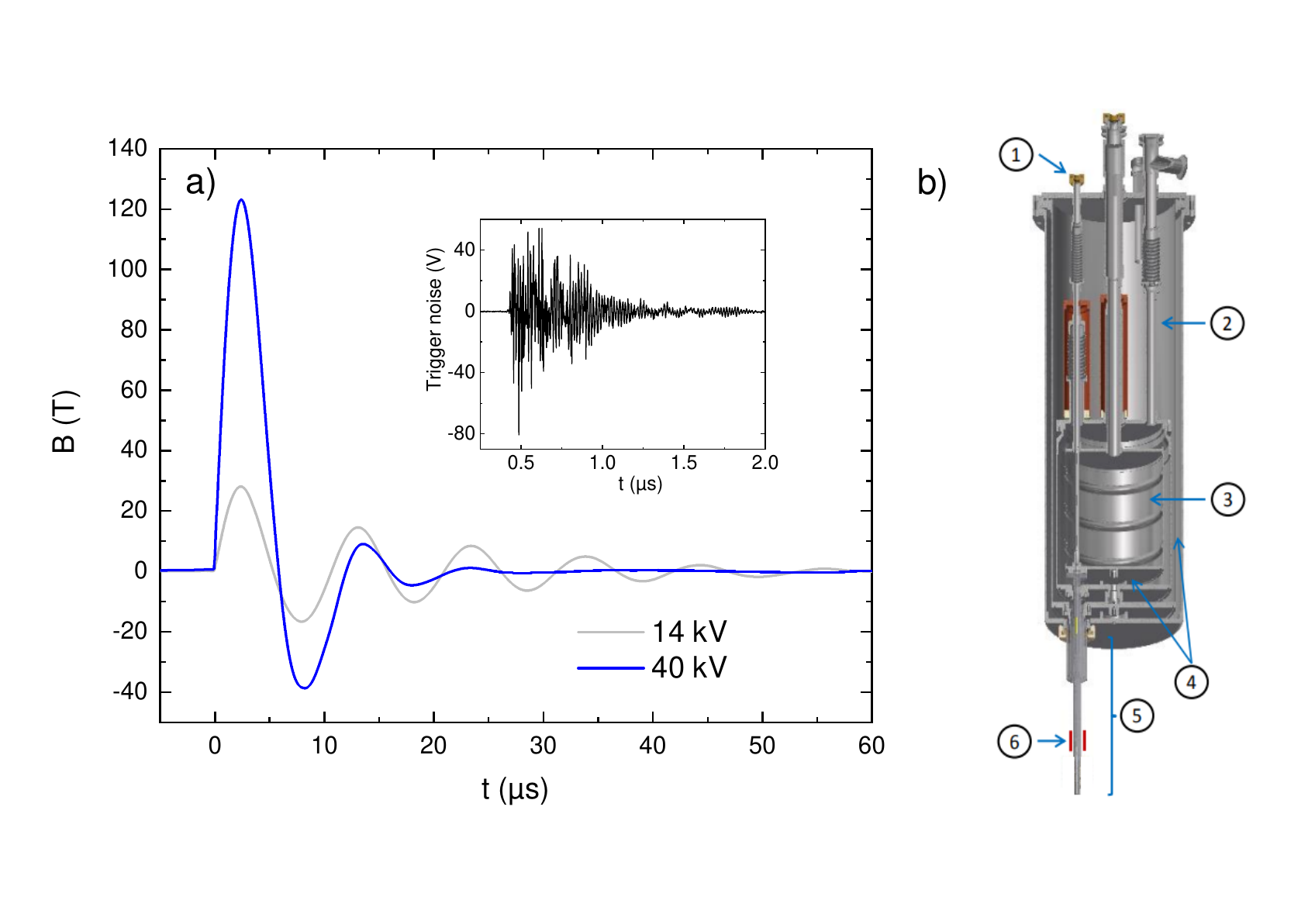}%
\caption{a) Magnetic field as a function of time for two different pulse modes; non-destructive in grey (14~kV) and semi-destructive in blue (40~kV). Inset: Measurement of the trigger noise coming from the spark-gap switches using a twisted pair soldered at its end and attenuators to protect the data acquisition card.
b) Sectional view of the cryostat showing (1) probe port, (2) nitrogen bath, (3) helium bath, (4) insulation vacuum, (5) polycarbonate extensions, and (6) single-turn coil. The experimental sample space has a diameter of 3.1~mm.
\label{Figure1}}%
\end{figure}

At the LNCMI, static magnetic field up to 42~T in Grenoble and non-destructive pulse magnetic fields up to 98~T in Toulouse can be generated.
To go beyond 100~T we have access to the megagauss installation in Toulouse \cite{Portugall1999}, where a capacitor bank is rapidly discharged into a single-turn copper coil, producing magnetic fields up to 310~T. 
Although the maximum magnetic field is lower than the one using flux compression technique, the controlled coil explosion enables repetitive and reproducible measurements in a safe sample space. For safety reason, the magnet cell is a large Faraday cage, and no galvanic connections are allowed outside of it.
In the megagauss installation, two pulse modes are available: a non-destructive one, where the single-coil is reinforced by a stain-steel clamp, with a maximum field of about 30~T, and a semi-destructive one with a maximum field of 150~T and above, depending on the diameter of the coil.
The installation produces microsecond pulse and the maximum field is reached in approximately 2 $\mu s$. Fig.~\ref{Figure1}~a) shows magnetic field versus time for a pulse in a reinforced mode (14~kV -- this value corresponds to the charging voltage of the capacitor bank) and for a semi-destructive pulse (40~kV).
The capacitor discharge is controlled using spark-gap switches whose triggering with a 50 kV - 1 ns pulse gives rise to extreme electromagnetic perturbations. Without efficient screening, these perturbations tend to affect electrical measurements during the up-sweep of the field thereby obscuring sensitive data as shown in the inset of Fig.~\ref{Figure1}~a). 

The setup is equipped with a metal–plastic cryostat consisting of a vacuum vessel that houses a helium tank, surrounded by a liquid nitrogen bath to minimize thermal radiation onto the helium reservoir (Fig.~\ref{Figure1}~b)). This configuration enables experiments down to pumped liquid helium temperatures (1.5~K). The cryostat extensions, made of polycarbonate tubes, are designed to fit within the 12 mm bore of the coil and are removable. This limits currently the maximum attainable magnetic field in our measurements to 150~T.

This article focuses on the development of transport measurements under these extreme conditions.
The first constraint to consider is the large $dB/dt$ exceeding $10^8~T/s$ at the beginning of the pulse. The induced voltage generated by such a large variation in the magnetic field are of the order of 100~V in a 1~mm$^2$ open loop.  Moreover, the noise from the spark-gap switches can damage the equipment. Finally, the large $dB/dt$ causes also Eddy current in metallic samples and self-heating.

\subsection{Experimental setup}
We have built a setup based on a two-point transmission measurement in the microwave range. We use a lock-in technique at $f \approx$ 1~GHz where the raw data are collected during the pulse and are demodulated afterwards using a digital software.
Data acquisition is carried out using the Teledyne SP devices card, ADQ7, with 14 bits resolution and 10 GS/s sampling rate. 

\begin{figure*}
\includegraphics[scale=0.25]{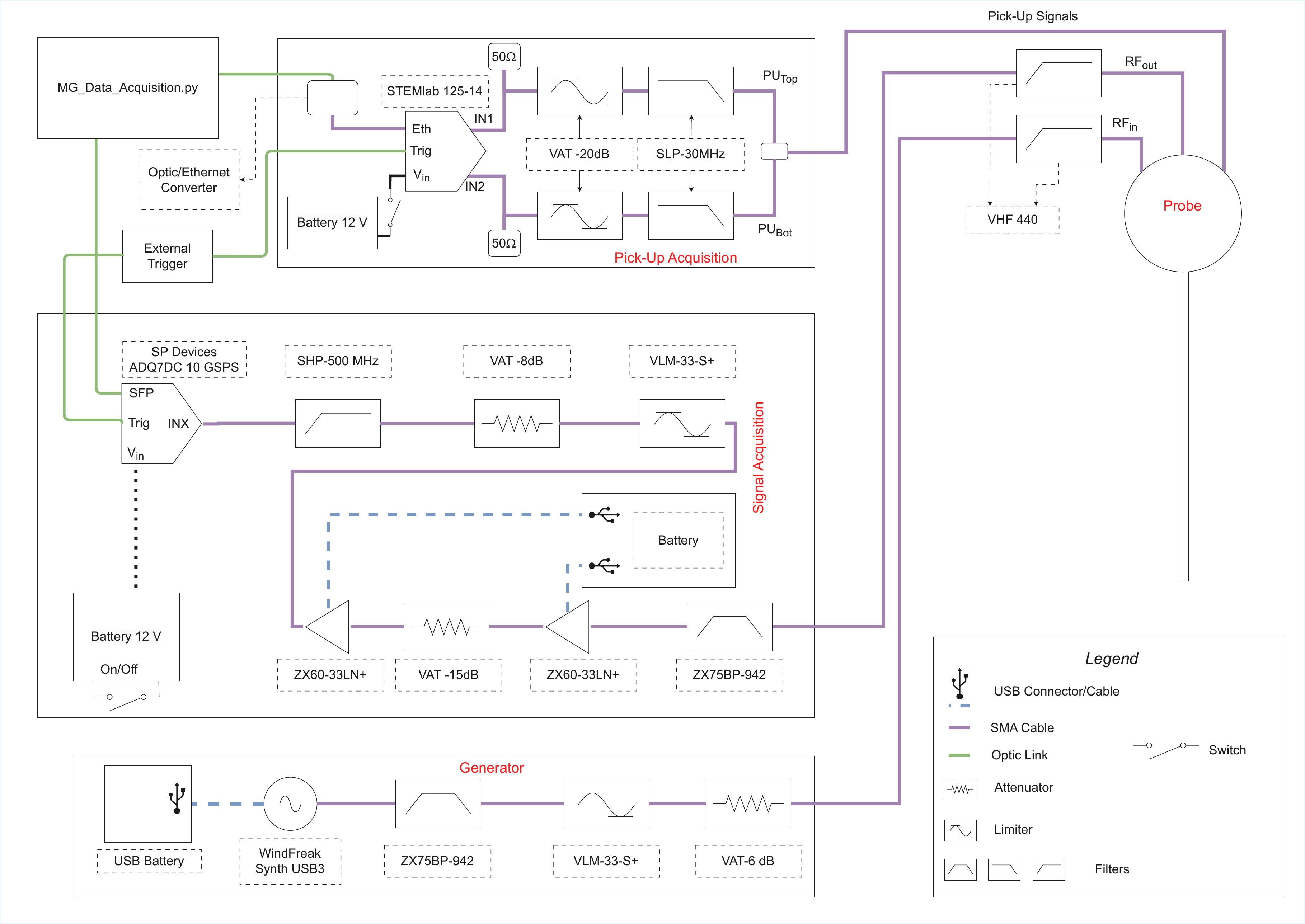}%
\caption{Block diagram of the microwave measurements. The circuit is separated in three main blocks. Top: pick-up acquisition card (RedPitaya Stemlab 125-14) with attenuator, filter and limiter. Middle: the signal acquisition data acquisition card (Teledyne SP devices card ADQ7) with filters, attenuators, pre-amplifiers and limiter. Bottom: the generator (a Windfreak synthUSB3) with  filter, attenuator and limiter. The three blocks are connected to the probe through double shielded coaxial cables. The correspondence between the symbols is shown in the legend box. \label{Figure2}}%
\end{figure*}

The complete block diagram for the microwave measurement is shown in Fig.~\ref{Figure2}.
From bottom to top, the schematics consist of three main blocks: the generator, the signal acquisition and the pick-up acquisition, respectively.
To generate the microwave signal we used the Windfreak synthUSB3 generator. It is protected from electromagnetic noise and reflection using band-pass filter, limiter, and attenuator.
The data acquisition card (ADQ7) is protected by filters, attenuators, limiter. The excellent signal-to-noise ratio is achieved by using two low-noise amplifiers.
A separated system is used for the pick-up acquisition, which is a RedPitaya Stemlab 125-14 with 14 bits resolution and 125 MS/S acquisition rate.
To protect the equipment from the spark gap noise and explosion, small Faraday cages were constructed, and all devices were placed inside, each powered by its own battery. Optical fibers are used to establish connections to the outside of the magnet cell.

The probe consists of small coaxial cables that terminate in twisted pairs for the connection to the sample. The pairs are shielded with cupro-nickel capillary tubes down to 1 mm above the center of the magnetic field. Two pick-up coils, made of two turns of copper wire wound around a 400~$\mu m$ ferrule, are positioned on opposite sides of the sample. At the output of the probe, band pass filters are connected to both coaxial cable with signal going in and out to protect the equipment from the large induced voltage.

\subsection{Heating model}

To estimate the heating of a sample subject to rapid magnetic field variations over a short period of time, we employ an adiabatic heating model. 
The sample is modeled as a stationary cylinder, whose axis is collinear with the applied magnetic field, and whose transverse surface matches that of the actual sample. 
The induced currents generated within the sample lead to Joule dissipation and thus to an internal temperature rise. 
This temperature gradient is proportional to the square of the time integral of the magnetic field derivative and is given by: 

\begin{equation}
 \Delta T = 
 \frac{r^2}{8~D}\int_{t_0}^{t_f}\frac{1}{\rho(t)~C_{v}(t)}\left(\frac{dB}{dt}\right)^2dt
 \label{Equation1}%
\end{equation}

where $r$ is the equivalent radius of the sample, $D$, $\rho$ and $C_{v}$ represent the density, the resistivity and the specific heat of the sample, respectively. The field dependence of the resistivity and the temperature dependence of the specific heat are taken into account in the calculation.

\section{Results}
Prior to the semi-destructive measurements, we performed 2-points transport measurements of the same sample in conventional pulsed fields at the LNCMI-Toulouse using a high input impedance acquisition card. The field dependence of the resistance $R$ of the sample is measured to calibrate the semi-destructive experiments. 
In the megagauss installation, the electronic circuit (Fig.~\ref{Figure2}) has 50~$\Omega$ impedance matching. The amplitude ($V$) of the signal obtained from the lock-in technique is converted to a resistance using the simple formula $V=\frac{50 e}{100+R}$, where $e$ is the effective voltage applied to the sample, taking into account the losses from the circuit and the probe. This parameter is adjusted using the absolute resistance of the sample measured in conventional pulsed fields.

\subsection{Metal Insulator Transition and Quantum Limit in InAs}

\begin{figure}
\includegraphics[scale=0.35]{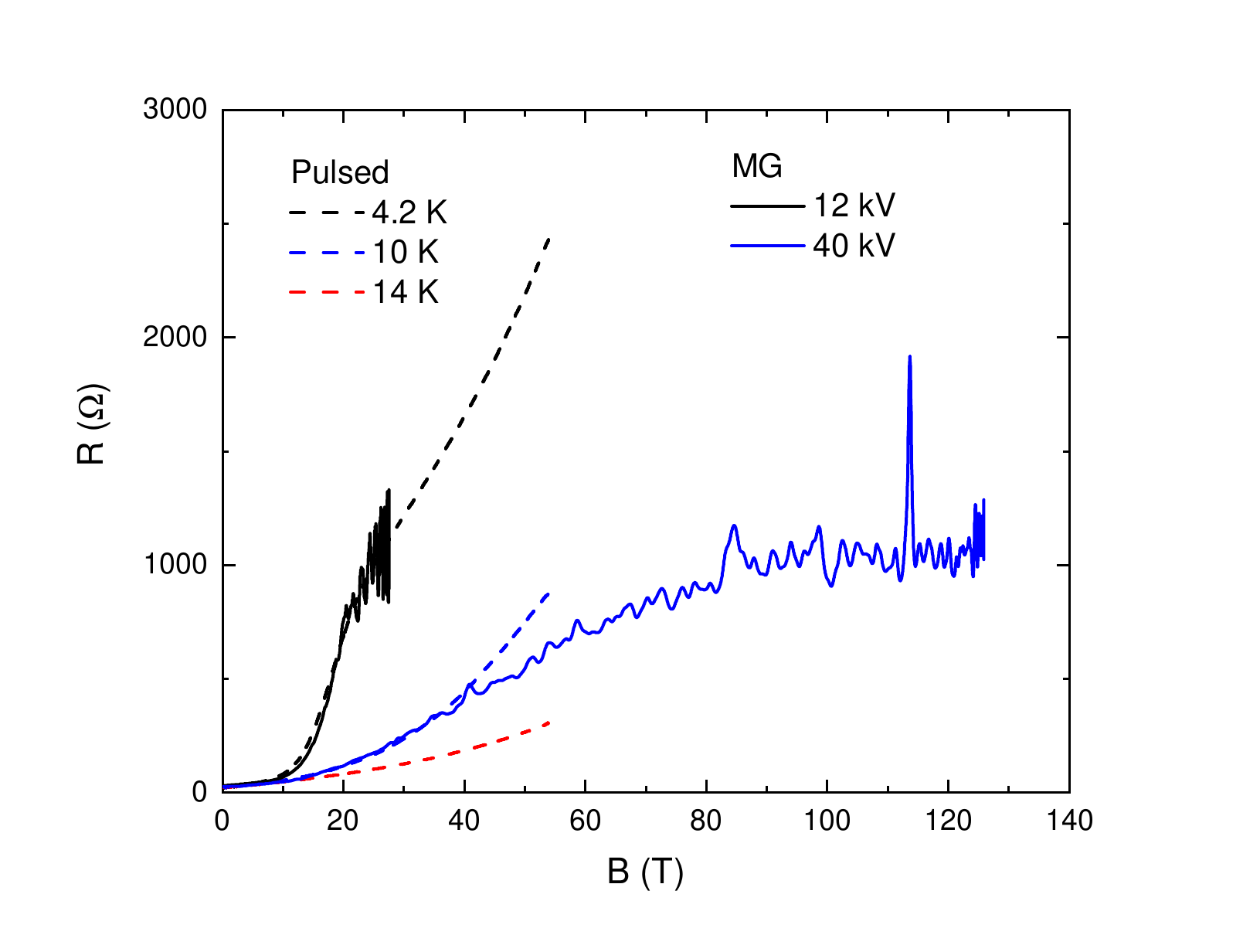}%
\caption{A comparison of the resistance of InAs measured in conventional pulsed magnetic fields (dashed lines) at $T$=4.2~K (black), 10~K (blue), and 14~K (red)  with that measured in the megagauss (MG) installation (solid lines) at 14~kV (black) and 40~kV (semi-destructive in blue). The base temperature before the pulse was $T$=4.2~K. The non-destructive megagauss data at 14~kV matches closely with the conventional pulsed field measurement at $T$=4.2~K. During the semi-destructive experiment at 40~kV (blue), the sample's temperature rises at the beginning of the pulse.  \label{Figure3}}%
\end{figure}

The first sample of interest is InAs, a narrow-gap semiconductor. The sample is lightly doped, with a carrier density $n = 1.6$~x~$10^{16}~cm^{-3}$. Above the quantum limit, that occurs at $B_{QL}$=4.1~T, when all carriers are confined in the lowest Landau level, it has been shown that a magnetic field assisted Mott-Anderson metal insulator (MI) transition occurs around 10~T at $T$=4.2~K \cite{Jaoui2020}.
The significant variation in the sample’s resistance at the MI transition with magnetic field and temperature served as an ideal benchmark for testing our setup. 
Figure \ref{Figure3} shows a comparison of the measurements in conventional pulsed fields (dashed lines) with the ones in the megagauss installation (solid line). In black ($T$=4.2~K), the megagauss data (12 kV) corresponding to a non-destructive pulse match perfectly with the conventional pulsed field data. A strong increase of the resistance is observed above 10~T, in good agreement with published data\cite{Jaoui2020}. For the semi-destructive pulse (40~kV in blue), the MI transition is shifted to higher magnetic fields, indicating that our sample is heating at the beginning of the pulse. Indeed, a comparison with adiabatic pulsed-field measurements indicates a temperature increase of approximately $\Delta T\approx$ 6~K. Nevertheless, this successful test validates our setup, and in particular, confirms that its sensitivity is sufficient to observe the metal-insulator transition.

\begin{figure*}
\includegraphics[scale=0.65]{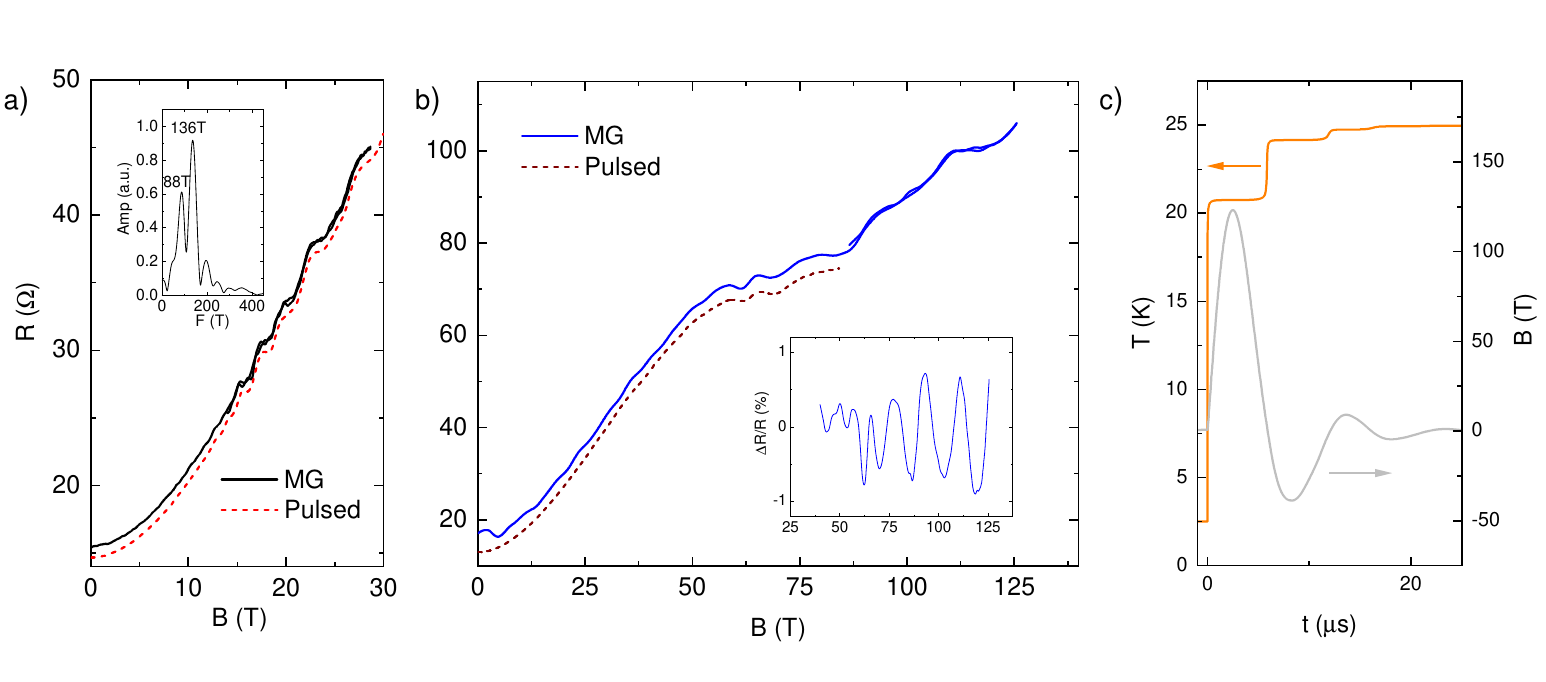}%
\caption{ Resistance of WTe$_{2}$ as a function of magnetic field. 
a) Comparison between non-destructive megagauss (MG) experiment (solid black line) and conventional pulse magnetic field (dashed red line) at $T$=4.2~K. The data are shifted for clarity.  Inset: Fourier transform of the quantum oscillations observed in megagauss. The observed frequencies are in excellent agreement with published data \cite{Linnartz2022}. 
b) Measurement in the semi-destructive mode at 40~kV (solid blue line) showing quantum oscillations up to 125~T. The base temperature before the pulse was $T$=2.5~K. The dashed line corresponds to measurement in the conventional pulsed field at $T$=20~K for comparison (the data are shifted for clarity). Note the reproducibility of the data at 40~kV during the rise above 95 T and the fall of the field pulse. Data for the rise of the field below 95~T are noisy because of the trigger noise. Inset: Quantum oscillations as a function of magnetic field with the spline background subtracted.
c) Time evolution of the temperature calculated from equation~\ref{Equation1} during a semi-destrutive pulse up to 125~T (orange line). The grey line shows the corresponding time dependence of the magnetic field. The temperature rises immediately at the beginning of the pulse, when both d$B$/d$t$ reaches its maximum and the resistivity is at a minimum. It then exhibits a plateau at $T$=21~K during the rise and the fall of the magnetic field (see also ref.~\onlinecite{Zimmerer2019}). In this regime, dissipation remains low due to the large magnetoresistance of WTe$_2$.
\label{Figure4}}%
\end{figure*}

\subsection{Quantum Oscillations in WTe$_{2}$}
Building on the encouraging results obtained in InAs, we moved on to detecting quantum oscillations, that allow to probe the topology of the Fermi surface of metals (FS). Indeed, the frequency of the oscillation (periodic in 1/$B$) is directly proportional to the area of the FS perpendicular to the magnetic field\cite{Shoenberg}. 
However, it is challenging to observe this phenomenon since the amplitude of the oscillations strongly depends on the temperature and is often small compared to the magnetoresistance. 
A good signal to noise ratio is therefore needed to observe this effect.
Here, we measured the Weyl semi-metal WTe$_{2}$. Its Fermi surface is composed of two pairs of electron-like pockets and two pairs of hole-like pockets, corresponding to quantum oscillation frequencies ranging from $F$=88~T to $F$=158~T \cite{Linnartz2022}. The pockets are in the nested Russian-doll arrangement, allowing the observation of magnetic breakdown between pockets of the same sign, e.g. frequency combination.

Fig.~\ref{Figure4}a compares measurements of the same sample of WTe$_{2}$ in the megagauss installation (14~kV black, non-destructive mode) with data in the conventional pulsed field (dashed red). 
Data are shifted for clarity.
The agreement between the two measurements is excellent, demonstrating that we can resolve quantum oscillations in the megagauss installation, despite the large d$B$/d$t$.
The inset displays the Fourier transform of the oscillatory part (background subtracted) of the megagauss data at 14~kV. 
The observed frequencies are in excellent agreement with previously reported data \cite{Linnartz2022}.
Fig.~ \ref{Figure4}b compares measurements in WTe$_2$ taken in the megagauss installation up to 125~T (semi-destructive mode 40~kV in blue) with data in the conventional pulsed field up to 86~T at $T$=20~K (dashed wine).
Note the reproducibility of the data at 40~kV during the rise above 95 T and the fall of the field pulse, which indicate a negligible heating in this field range.
In addition, quantum oscillations observed up to 86~T are in excellent agreement with measurements in the conventional pulsed field installation. However, a comparison of the amplitude of quantum oscillations allows to estimate heating of the sample at the beginning of the pulse to be about $\Delta T \approx$~17~K. Fig.~\ref{Figure4}c shows the temperature variation of our sample during a semi-destructive pulse using Equation~\ref{Equation1} and published data for specific heat \cite{Callanan92} and resistivity \cite{Jha18}.
The heating arises at the beginning of the pulse and equation~\ref{Equation1} leads to $\Delta T \approx$~18~K, in good agreement with our estimation. Then, there is a plateau in the temperature during the rise and the fall of the magnetic field.
Despite the heating of the sample and due to the light effective mass of WTe$_2$, our setup enabled the observation of quantum oscillations in the megagauss installation for a semi-destructive pulse. In the inset of Fig.~\ref{Figure4}b, the oscillatory part of quantum oscillations up to $B$ = 125~T are shown as a function of magnetic field.
From a plot of position of maxima and minima of quantum oscillations, we estimate frequencies between approximately 400~T and 700~T.
They originate from magnetic breakdown between the four main pockets and the 700~T oscillation was never observed before.

\section{Conclusion}
In this paper we have reported the development of a transport measurement with excellent signal to noise in the megagauss installation.
We reported the observation of a metal insulator transition in the quantum limit of InAs.
We also reported the observation of quantum oscillations in the Weyl semi-metal WTe$_{2}$, leading to the first observation of quantum oscillations in a megagauss experiment above 100~T. To reduce heating of the sample, work is in progress to reduce their size, in particular using the focusing ion beam technique. Our work paves the way for transport properties in quantum materials at fields up to 200~T in the megagauss facility.\\

\begin{acknowledgments}
 We thank G. Rikken for helpful and stimulating discussions and Mathieu Barragan for his help to build the cryostat. D.V. and C.P. acknowledge support from the EUR grant NanoX no ANR-17-EURE-0009 and from the ANR grant NEPTUN no ANR-19-CE30-0019-01. 
 This work was supported by LNCMI-CNRS, members of the European Magnetic Field Laboratory (EMFL), the Canada Research Chair program, the NSERC and the Canada First Research Excellence Fund.
\end{acknowledgments}

\bibliography{MG_RSI_bib}

\begin{thebibliography}{20}%
\makeatletter
\providecommand \@ifxundefined [1]{%
 \@ifx{#1\undefined}
}%
\providecommand \@ifnum [1]{%
 \ifnum #1\expandafter \@firstoftwo
 \else \expandafter \@secondoftwo
 \fi
}%
\providecommand \@ifx [1]{%
 \ifx #1\expandafter \@firstoftwo
 \else \expandafter \@secondoftwo
 \fi
}%
\providecommand \natexlab [1]{#1}%
\providecommand \enquote  [1]{``#1''}%
\providecommand \bibnamefont  [1]{#1}%
\providecommand \bibfnamefont [1]{#1}%
\providecommand \citenamefont [1]{#1}%
\providecommand \href@noop [0]{\@secondoftwo}%
\providecommand \href [0]{\begingroup \@sanitize@url \@href}%
\providecommand \@href[1]{\@@startlink{#1}\@@href}%
\providecommand \@@href[1]{\endgroup#1\@@endlink}%
\providecommand \@sanitize@url [0]{\catcode `\\12\catcode `\$12\catcode `\&12\catcode `\#12\catcode `\^12\catcode `\_12\catcode `\%12\relax}%
\providecommand \@@startlink[1]{}%
\providecommand \@@endlink[0]{}%
\providecommand \url  [0]{\begingroup\@sanitize@url \@url }%
\providecommand \@url [1]{\endgroup\@href {#1}{\urlprefix }}%
\providecommand \urlprefix  [0]{URL }%
\providecommand \Eprint [0]{\href }%
\providecommand \doibase [0]{http://dx.doi.org/}%
\providecommand \selectlanguage [0]{\@gobble}%
\providecommand \bibinfo  [0]{\@secondoftwo}%
\providecommand \bibfield  [0]{\@secondoftwo}%
\providecommand \translation [1]{[#1]}%
\providecommand \BibitemOpen [0]{}%
\providecommand \bibitemStop [0]{}%
\providecommand \bibitemNoStop [0]{.\EOS\space}%
\providecommand \EOS [0]{\spacefactor3000\relax}%
\providecommand \BibitemShut  [1]{\csname bibitem#1\endcsname}%
\let\auto@bib@innerbib\@empty
\bibitem [{\citenamefont {Shoenberg}(1984)}]{Shoenberg}%
  \BibitemOpen
  \bibfield  {author} {\bibinfo {author} {\bibfnamefont {D.}~\bibnamefont {Shoenberg}},\ }\href@noop {} {\emph {\bibinfo {title} {Magnetic Oscillations in Metals}}}\ (\bibinfo  {publisher} {Cambridge Univ. Press, Cambridge},\ \bibinfo {year} {1984})\BibitemShut {NoStop}%
\bibitem [{\citenamefont {Boebinger}\ \emph {et~al.}(1996)\citenamefont {Boebinger}, \citenamefont {Ando}, \citenamefont {Passner}, \citenamefont {Kimura}, \citenamefont {Okuya}, \citenamefont {Shimoyama}, \citenamefont {Kishio}, \citenamefont {Tamasaku}, \citenamefont {Ichikawa},\ and\ \citenamefont {Uchida}}]{Boebinger1996}%
  \BibitemOpen
  \bibfield  {author} {\bibinfo {author} {\bibfnamefont {G.~S.}\ \bibnamefont {Boebinger}}, \bibinfo {author} {\bibfnamefont {Y.}~\bibnamefont {Ando}}, \bibinfo {author} {\bibfnamefont {A.}~\bibnamefont {Passner}}, \bibinfo {author} {\bibfnamefont {T.}~\bibnamefont {Kimura}}, \bibinfo {author} {\bibfnamefont {M.}~\bibnamefont {Okuya}}, \bibinfo {author} {\bibfnamefont {J.}~\bibnamefont {Shimoyama}}, \bibinfo {author} {\bibfnamefont {K.}~\bibnamefont {Kishio}}, \bibinfo {author} {\bibfnamefont {K.}~\bibnamefont {Tamasaku}}, \bibinfo {author} {\bibfnamefont {N.}~\bibnamefont {Ichikawa}}, \ and\ \bibinfo {author} {\bibfnamefont {S.}~\bibnamefont {Uchida}},\ }\bibfield  {title} {\enquote {\bibinfo {title} {Insulator-to-metal crossover in the normal state of lsco near optimum doping},}\ }\href {\doibase 10.1103/PhysRevLett.77.5417} {\bibfield  {journal} {\bibinfo  {journal} {Phys. Rev. Lett.}\ }\textbf {\bibinfo {volume} {77}},\ \bibinfo {pages} {5417--5420} (\bibinfo {year} {1996})}\BibitemShut {NoStop}%
\bibitem [{\citenamefont {Cooper}\ \emph {et~al.}(2009)\citenamefont {Cooper}, \citenamefont {Wang}, \citenamefont {Vignolle}, \citenamefont {Lipscombe}, \citenamefont {Hayden}, \citenamefont {Tanabe}, \citenamefont {Adachi}, \citenamefont {Koike}, \citenamefont {Nohara}, \citenamefont {Takagi}, \citenamefont {Proust},\ and\ \citenamefont {Hussey}}]{Cooper2009}%
  \BibitemOpen
  \bibfield  {author} {\bibinfo {author} {\bibfnamefont {R.~A.}\ \bibnamefont {Cooper}}, \bibinfo {author} {\bibfnamefont {Y.}~\bibnamefont {Wang}}, \bibinfo {author} {\bibfnamefont {B.}~\bibnamefont {Vignolle}}, \bibinfo {author} {\bibfnamefont {O.~J.}\ \bibnamefont {Lipscombe}}, \bibinfo {author} {\bibfnamefont {S.~M.}\ \bibnamefont {Hayden}}, \bibinfo {author} {\bibfnamefont {Y.}~\bibnamefont {Tanabe}}, \bibinfo {author} {\bibfnamefont {T.}~\bibnamefont {Adachi}}, \bibinfo {author} {\bibfnamefont {Y.}~\bibnamefont {Koike}}, \bibinfo {author} {\bibfnamefont {M.}~\bibnamefont {Nohara}}, \bibinfo {author} {\bibfnamefont {H.}~\bibnamefont {Takagi}}, \bibinfo {author} {\bibfnamefont {C.}~\bibnamefont {Proust}}, \ and\ \bibinfo {author} {\bibfnamefont {N.~E.}\ \bibnamefont {Hussey}},\ }\bibfield  {title} {\enquote {\bibinfo {title} {Anomalous criticality in the electrical resistivity of lsco},}\ }\href {\doibase 10.1126/science.1165015} {\bibfield  {journal} {\bibinfo  {journal} {Science}\ }\textbf {\bibinfo
  {volume} {323}},\ \bibinfo {pages} {603--607} (\bibinfo {year} {2009})}\BibitemShut {NoStop}%
\bibitem [{\citenamefont {Doiron-Leyraud}\ \emph {et~al.}(2007)\citenamefont {Doiron-Leyraud}, \citenamefont {Proust}, \citenamefont {LeBoeuf}, \citenamefont {Levallois}, \citenamefont {Bonnemaison}, \citenamefont {Liang}, \citenamefont {Bonn}, \citenamefont {Hardy},\ and\ \citenamefont {Taillefer}}]{Doiron2007}%
  \BibitemOpen
  \bibfield  {author} {\bibinfo {author} {\bibfnamefont {N.}~\bibnamefont {Doiron-Leyraud}}, \bibinfo {author} {\bibfnamefont {C.}~\bibnamefont {Proust}}, \bibinfo {author} {\bibfnamefont {D.}~\bibnamefont {LeBoeuf}}, \bibinfo {author} {\bibfnamefont {J.}~\bibnamefont {Levallois}}, \bibinfo {author} {\bibfnamefont {J.-B.}\ \bibnamefont {Bonnemaison}}, \bibinfo {author} {\bibfnamefont {R.}~\bibnamefont {Liang}}, \bibinfo {author} {\bibfnamefont {D.}~\bibnamefont {Bonn}}, \bibinfo {author} {\bibfnamefont {W.}~\bibnamefont {Hardy}}, \ and\ \bibinfo {author} {\bibfnamefont {L.}~\bibnamefont {Taillefer}},\ }\bibfield  {title} {\enquote {\bibinfo {title} {Quantum oscillations and the fermi surface in an underdoped high-t$_c$ superconductor},}\ }\href {https://www.nature.com/articles/nature05872} {\bibfield  {journal} {\bibinfo  {journal} {Nature}\ }\textbf {\bibinfo {volume} {447}},\ \bibinfo {pages} {565--568} (\bibinfo {year} {2007})}\BibitemShut {NoStop}%
\bibitem [{\citenamefont {Stormer}, \citenamefont {Tsui},\ and\ \citenamefont {Gossard}(1999)}]{Stormer1999}%
  \BibitemOpen
  \bibfield  {author} {\bibinfo {author} {\bibfnamefont {H.~L.}\ \bibnamefont {Stormer}}, \bibinfo {author} {\bibfnamefont {D.~C.}\ \bibnamefont {Tsui}}, \ and\ \bibinfo {author} {\bibfnamefont {A.~C.}\ \bibnamefont {Gossard}},\ }\bibfield  {title} {\enquote {\bibinfo {title} {The fractional quantum hall effect},}\ }\href {\doibase 10.1103/RevModPhys.71.S298} {\bibfield  {journal} {\bibinfo  {journal} {Rev. Mod. Phys.}\ }\textbf {\bibinfo {volume} {71}},\ \bibinfo {pages} {S298--S305} (\bibinfo {year} {1999})}\BibitemShut {NoStop}%
\bibitem [{\citenamefont {Aoki}, \citenamefont {Knafo},\ and\ \citenamefont {Sheikin}(2013)}]{Aoki2013}%
  \BibitemOpen
  \bibfield  {author} {\bibinfo {author} {\bibfnamefont {D.}~\bibnamefont {Aoki}}, \bibinfo {author} {\bibfnamefont {W.}~\bibnamefont {Knafo}}, \ and\ \bibinfo {author} {\bibfnamefont {I.}~\bibnamefont {Sheikin}},\ }\bibfield  {title} {\enquote {\bibinfo {title} {Heavy fermions in a high magnetic field},}\ }\href {\doibase 10.1016/j.crhy.2012.11.004} {\bibfield  {journal} {\bibinfo  {journal} {Comptes Rendus. Physique}\ }\textbf {\bibinfo {volume} {14}},\ \bibinfo {pages} {53--77} (\bibinfo {year} {2013})}\BibitemShut {NoStop}%
\bibitem [{\citenamefont {Fauqu\'e}\ \emph {et~al.}(2013)\citenamefont {Fauqu\'e}, \citenamefont {LeBoeuf}, \citenamefont {Vignolle}, \citenamefont {Nardone}, \citenamefont {Proust},\ and\ \citenamefont {Behnia}}]{Fauque2013}%
  \BibitemOpen
  \bibfield  {author} {\bibinfo {author} {\bibfnamefont {B.}~\bibnamefont {Fauqu\'e}}, \bibinfo {author} {\bibfnamefont {D.}~\bibnamefont {LeBoeuf}}, \bibinfo {author} {\bibfnamefont {B.}~\bibnamefont {Vignolle}}, \bibinfo {author} {\bibfnamefont {M.}~\bibnamefont {Nardone}}, \bibinfo {author} {\bibfnamefont {C.}~\bibnamefont {Proust}}, \ and\ \bibinfo {author} {\bibfnamefont {K.}~\bibnamefont {Behnia}},\ }\bibfield  {title} {\enquote {\bibinfo {title} {Two phase transitions induced by a magnetic field in graphite},}\ }\href {\doibase 10.1103/PhysRevLett.110.266601} {\bibfield  {journal} {\bibinfo  {journal} {Phys. Rev. Lett.}\ }\textbf {\bibinfo {volume} {110}},\ \bibinfo {pages} {266601} (\bibinfo {year} {2013})}\BibitemShut {NoStop}%
\bibitem [{\citenamefont {Nguyen}, \citenamefont {Michel},\ and\ \citenamefont {Mielke}(2016)}]{Nguyen2016}%
  \BibitemOpen
  \bibfield  {author} {\bibinfo {author} {\bibfnamefont {D.~N.}\ \bibnamefont {Nguyen}}, \bibinfo {author} {\bibfnamefont {J.}~\bibnamefont {Michel}}, \ and\ \bibinfo {author} {\bibfnamefont {C.~H.}\ \bibnamefont {Mielke}},\ }\bibfield  {title} {\enquote {\bibinfo {title} {Status and development of pulsed magnets at the nhmfl pulsed field facility},}\ }\href {\doibase 10.1109/TASC.2016.2515982} {\bibfield  {journal} {\bibinfo  {journal} {IEEE Transactions on Applied Superconductivity}\ }\textbf {\bibinfo {volume} {26}},\ \bibinfo {pages} {1--5} (\bibinfo {year} {2016})}\BibitemShut {NoStop}%
\bibitem [{\citenamefont {Béard}\ \emph {et~al.}(2018)\citenamefont {Béard}, \citenamefont {Billette}, \citenamefont {Ferreira}, \citenamefont {Frings}, \citenamefont {Lagarrigue}, \citenamefont {Lecouturier},\ and\ \citenamefont {Nicolin}}]{Beard2018}%
  \BibitemOpen
  \bibfield  {author} {\bibinfo {author} {\bibfnamefont {J.}~\bibnamefont {Béard}}, \bibinfo {author} {\bibfnamefont {J.}~\bibnamefont {Billette}}, \bibinfo {author} {\bibfnamefont {N.}~\bibnamefont {Ferreira}}, \bibinfo {author} {\bibfnamefont {P.}~\bibnamefont {Frings}}, \bibinfo {author} {\bibfnamefont {J.-M.}\ \bibnamefont {Lagarrigue}}, \bibinfo {author} {\bibfnamefont {F.}~\bibnamefont {Lecouturier}}, \ and\ \bibinfo {author} {\bibfnamefont {J.-P.}\ \bibnamefont {Nicolin}},\ }\bibfield  {title} {\enquote {\bibinfo {title} {Design and tests of the 100-t triple coil at lncmi},}\ }\href {\doibase 10.1109/TASC.2017.2779817} {\bibfield  {journal} {\bibinfo  {journal} {IEEE Transactions on Applied Superconductivity}\ }\textbf {\bibinfo {volume} {28}},\ \bibinfo {pages} {1--5} (\bibinfo {year} {2018})}\BibitemShut {NoStop}%
\bibitem [{\citenamefont {Miura}\ \emph {et~al.}(2002)\citenamefont {Miura}, \citenamefont {Nakagawa}, \citenamefont {Sekitani}, \citenamefont {Naito}, \citenamefont {Sato},\ and\ \citenamefont {Enomoto}}]{Miura2002}%
  \BibitemOpen
  \bibfield  {author} {\bibinfo {author} {\bibfnamefont {N.}~\bibnamefont {Miura}}, \bibinfo {author} {\bibfnamefont {H.}~\bibnamefont {Nakagawa}}, \bibinfo {author} {\bibfnamefont {T.}~\bibnamefont {Sekitani}}, \bibinfo {author} {\bibfnamefont {M.}~\bibnamefont {Naito}}, \bibinfo {author} {\bibfnamefont {H.}~\bibnamefont {Sato}}, \ and\ \bibinfo {author} {\bibfnamefont {Y.}~\bibnamefont {Enomoto}},\ }\bibfield  {title} {\enquote {\bibinfo {title} {High-magnetic-field study of high-t$_c$ cuprates},}\ }\href {\doibase https://doi.org/10.1016/S0921-4526(02)01134-1} {\bibfield  {journal} {\bibinfo  {journal} {Physica B: Condensed Matter}\ }\textbf {\bibinfo {volume} {319}},\ \bibinfo {pages} {310--320} (\bibinfo {year} {2002})}\BibitemShut {NoStop}%
\bibitem [{\citenamefont {Shimamoto}, \citenamefont {Miura},\ and\ \citenamefont {Nojiri}(1998)}]{Shimamoto1998}%
  \BibitemOpen
  \bibfield  {author} {\bibinfo {author} {\bibfnamefont {Y.}~\bibnamefont {Shimamoto}}, \bibinfo {author} {\bibfnamefont {N.}~\bibnamefont {Miura}}, \ and\ \bibinfo {author} {\bibfnamefont {H.}~\bibnamefont {Nojiri}},\ }\bibfield  {title} {\enquote {\bibinfo {title} {Magnetic-field-induced electronic phase transitions in semimetals in high magnetic fields},}\ }\href {\doibase 10.1088/0953-8984/10/49/018} {\bibfield  {journal} {\bibinfo  {journal} {Journal of Physics: Condensed Matter}\ }\textbf {\bibinfo {volume} {10}},\ \bibinfo {pages} {11289} (\bibinfo {year} {1998})}\BibitemShut {NoStop}%
\bibitem [{\citenamefont {Sekitani}, \citenamefont {Matsuda},\ and\ \citenamefont {Miura}(2007)}]{Sekitani2007}%
  \BibitemOpen
  \bibfield  {author} {\bibinfo {author} {\bibfnamefont {T.}~\bibnamefont {Sekitani}}, \bibinfo {author} {\bibfnamefont {Y.~H.}\ \bibnamefont {Matsuda}}, \ and\ \bibinfo {author} {\bibfnamefont {N.}~\bibnamefont {Miura}},\ }\bibfield  {title} {\enquote {\bibinfo {title} {Measurement of the upper critical field of optimally-doped ybco in megagauss magnetic fields},}\ }\href {\doibase 10.1088/1367-2630/9/3/047} {\bibfield  {journal} {\bibinfo  {journal} {New Journal of Physics}\ }\textbf {\bibinfo {volume} {9}},\ \bibinfo {pages} {47} (\bibinfo {year} {2007})}\BibitemShut {NoStop}%
\bibitem [{\citenamefont {Nakamura}, \citenamefont {Altarawneh},\ and\ \citenamefont {Takeyama}(2018)}]{Nakamura2018}%
  \BibitemOpen
  \bibfield  {author} {\bibinfo {author} {\bibfnamefont {D.}~\bibnamefont {Nakamura}}, \bibinfo {author} {\bibfnamefont {M.~M.}\ \bibnamefont {Altarawneh}}, \ and\ \bibinfo {author} {\bibfnamefont {S.}~\bibnamefont {Takeyama}},\ }\bibfield  {title} {\enquote {\bibinfo {title} {Radio frequency self-resonant coil for contactless ac-conductivity in 100 t class ultra-strong pulse magnetic fields},}\ }\href {\doibase 10.1088/1361-6501/aa9a0b} {\bibfield  {journal} {\bibinfo  {journal} {Measurement Science and Technology}\ }\textbf {\bibinfo {volume} {29}},\ \bibinfo {pages} {035901} (\bibinfo {year} {2018})}\BibitemShut {NoStop}%
\bibitem [{\citenamefont {Shitaokoshi}\ \emph {et~al.}(2023)\citenamefont {Shitaokoshi}, \citenamefont {Kawachi}, \citenamefont {Nomura}, \citenamefont {Balakirev},\ and\ \citenamefont {Kohama}}]{Shitaokoshi2023}%
  \BibitemOpen
  \bibfield  {author} {\bibinfo {author} {\bibfnamefont {T.}~\bibnamefont {Shitaokoshi}}, \bibinfo {author} {\bibfnamefont {S.}~\bibnamefont {Kawachi}}, \bibinfo {author} {\bibfnamefont {T.}~\bibnamefont {Nomura}}, \bibinfo {author} {\bibfnamefont {F.~F.}\ \bibnamefont {Balakirev}}, \ and\ \bibinfo {author} {\bibfnamefont {Y.}~\bibnamefont {Kohama}},\ }\bibfield  {title} {\enquote {\bibinfo {title} {Radio frequency electrical resistance measurement under destructive pulsed magnetic fields},}\ }\href {\doibase 10.1063/5.0165680} {\bibfield  {journal} {\bibinfo  {journal} {Review of Scientific Instruments}\ }\textbf {\bibinfo {volume} {94}},\ \bibinfo {pages} {094706} (\bibinfo {year} {2023})}\BibitemShut {NoStop}%
\bibitem [{\citenamefont {Jaoui}\ \emph {et~al.}(2020)\citenamefont {Jaoui}, \citenamefont {Seyfarth}, \citenamefont {Rischau}, \citenamefont {Wiedmann}, \citenamefont {Benhabib}, \citenamefont {Proust}, \citenamefont {Behnia},\ and\ \citenamefont {Fauqué}}]{Jaoui2020}%
  \BibitemOpen
  \bibfield  {author} {\bibinfo {author} {\bibfnamefont {A.}~\bibnamefont {Jaoui}}, \bibinfo {author} {\bibfnamefont {G.}~\bibnamefont {Seyfarth}}, \bibinfo {author} {\bibfnamefont {C.~W.}\ \bibnamefont {Rischau}}, \bibinfo {author} {\bibfnamefont {S.}~\bibnamefont {Wiedmann}}, \bibinfo {author} {\bibfnamefont {S.}~\bibnamefont {Benhabib}}, \bibinfo {author} {\bibfnamefont {C.}~\bibnamefont {Proust}}, \bibinfo {author} {\bibfnamefont {K.}~\bibnamefont {Behnia}}, \ and\ \bibinfo {author} {\bibfnamefont {B.}~\bibnamefont {Fauqué}},\ }\href@noop {} {\bibfield  {journal} {\bibinfo  {journal} {npj Quantum Materials}\ }\textbf {\bibinfo {volume} {5}},\ \bibinfo {pages} {94} (\bibinfo {year} {2020})}\BibitemShut {NoStop}%
\bibitem [{\citenamefont {Linnartz}\ \emph {et~al.}(2022)\citenamefont {Linnartz}, \citenamefont {Müller}, \citenamefont {Hsu}, \citenamefont {Nielsen}, \citenamefont {Bremholm}, \citenamefont {Hussey}, \citenamefont {Carrington},\ and\ \citenamefont {Wiedmann}}]{Linnartz2022}%
  \BibitemOpen
  \bibfield  {author} {\bibinfo {author} {\bibfnamefont {J.~F.}\ \bibnamefont {Linnartz}}, \bibinfo {author} {\bibfnamefont {C.~S.~A.}\ \bibnamefont {Müller}}, \bibinfo {author} {\bibfnamefont {Y.-T.}\ \bibnamefont {Hsu}}, \bibinfo {author} {\bibfnamefont {C.~B.}\ \bibnamefont {Nielsen}}, \bibinfo {author} {\bibfnamefont {M.}~\bibnamefont {Bremholm}}, \bibinfo {author} {\bibfnamefont {N.~E.}\ \bibnamefont {Hussey}}, \bibinfo {author} {\bibfnamefont {A.}~\bibnamefont {Carrington}}, \ and\ \bibinfo {author} {\bibfnamefont {S.}~\bibnamefont {Wiedmann}},\ }\href@noop {} {\bibfield  {journal} {\bibinfo  {journal} {Physical Review Research}\ }\textbf {\bibinfo {volume} {4}},\ \bibinfo {pages} {L012005} (\bibinfo {year} {2022})}\BibitemShut {NoStop}%
\bibitem [{\citenamefont {Portugall}\ \emph {et~al.}(1999)\citenamefont {Portugall}, \citenamefont {Puhlmann}, \citenamefont {Müller}, \citenamefont {Barczewski}, \citenamefont {Stolpe},\ and\ \citenamefont {von Ortenberg}}]{Portugall1999}%
  \BibitemOpen
  \bibfield  {author} {\bibinfo {author} {\bibfnamefont {O.}~\bibnamefont {Portugall}}, \bibinfo {author} {\bibfnamefont {N.}~\bibnamefont {Puhlmann}}, \bibinfo {author} {\bibfnamefont {H.}~\bibnamefont {Müller}}, \bibinfo {author} {\bibfnamefont {M.}~\bibnamefont {Barczewski}}, \bibinfo {author} {\bibfnamefont {I.}~\bibnamefont {Stolpe}}, \ and\ \bibinfo {author} {\bibfnamefont {M.}~\bibnamefont {von Ortenberg}},\ }\href@noop {} {\bibfield  {journal} {\bibinfo  {journal} {J. Phys. D: Appl. Phys.}\ }\textbf {\bibinfo {volume} {32}},\ \bibinfo {pages} {2354} (\bibinfo {year} {1999})}\BibitemShut {NoStop}%
\bibitem [{\citenamefont {Zimmerer}\ \emph {et~al.}(2019)\citenamefont {Zimmerer}, \citenamefont {Mejia}, \citenamefont {Utech}, \citenamefont {Arnhold}, \citenamefont {Janke},\ and\ \citenamefont {Wosnitza}}]{Zimmerer2019}%
  \BibitemOpen
  \bibfield  {author} {\bibinfo {author} {\bibfnamefont {C.}~\bibnamefont {Zimmerer}}, \bibinfo {author} {\bibfnamefont {C.~S.}\ \bibnamefont {Mejia}}, \bibinfo {author} {\bibfnamefont {T.}~\bibnamefont {Utech}}, \bibinfo {author} {\bibfnamefont {K.}~\bibnamefont {Arnhold}}, \bibinfo {author} {\bibfnamefont {A.}~\bibnamefont {Janke}}, \ and\ \bibinfo {author} {\bibfnamefont {J.}~\bibnamefont {Wosnitza}},\ }\bibfield  {title} {\enquote {\bibinfo {title} {Inductive heating using a high-magnetic-field pulse to initiate chemical reactions to generate composite materials},}\ }\href {\doibase 10.3390/polym11030535} {\bibfield  {journal} {\bibinfo  {journal} {Polymers}\ }\textbf {\bibinfo {volume} {11}},\ \bibinfo {pages} {535} (\bibinfo {year} {2019})}\BibitemShut {NoStop}%
\bibitem [{\citenamefont {Callanan}\ \emph {et~al.}(1992)\citenamefont {Callanan}, \citenamefont {Hope}, \citenamefont {Weir},\ and\ \citenamefont {Westrum}}]{Callanan92}%
  \BibitemOpen
  \bibfield  {author} {\bibinfo {author} {\bibfnamefont {J.~E.}\ \bibnamefont {Callanan}}, \bibinfo {author} {\bibfnamefont {G.}~\bibnamefont {Hope}}, \bibinfo {author} {\bibfnamefont {R.~D.}\ \bibnamefont {Weir}}, \ and\ \bibinfo {author} {\bibfnamefont {E.~F.}\ \bibnamefont {Westrum}},\ }\bibfield  {title} {\enquote {\bibinfo {title} {Thermodynamic properties of tungsten ditelluride (wte2) i. the preparation and lowtemperature heat capacity at temperatures from 6 k to 326 k},}\ }\href {\doibase https://doi.org/10.1016/S0021-9614(05)80034-5} {\bibfield  {journal} {\bibinfo  {journal} {The Journal of Chemical Thermodynamics}\ }\textbf {\bibinfo {volume} {24}},\ \bibinfo {pages} {627--638} (\bibinfo {year} {1992})}\BibitemShut {NoStop}%
\bibitem [{\citenamefont {Jha}\ \emph {et~al.}(2018)\citenamefont {Jha}, \citenamefont {Onishi}, \citenamefont {Higashinaka}, \citenamefont {Matsuda}, \citenamefont {Ribeiro},\ and\ \citenamefont {Aoki}}]{Jha18}%
  \BibitemOpen
  \bibfield  {author} {\bibinfo {author} {\bibfnamefont {R.}~\bibnamefont {Jha}}, \bibinfo {author} {\bibfnamefont {S.}~\bibnamefont {Onishi}}, \bibinfo {author} {\bibfnamefont {R.}~\bibnamefont {Higashinaka}}, \bibinfo {author} {\bibfnamefont {T.~D.}\ \bibnamefont {Matsuda}}, \bibinfo {author} {\bibfnamefont {R.~A.}\ \bibnamefont {Ribeiro}}, \ and\ \bibinfo {author} {\bibfnamefont {Y.}~\bibnamefont {Aoki}},\ }\bibfield  {title} {\enquote {\bibinfo {title} {Anisotropy in the electronic transport properties of weyl semimetal wte2 single crystals},}\ }\href {\doibase 10.1063/1.5043063} {\bibfield  {journal} {\bibinfo  {journal} {AIP Advances}\ }\textbf {\bibinfo {volume} {8}},\ \bibinfo {pages} {101332} (\bibinfo {year} {2018})}\BibitemShut {NoStop}%
\end{thebibliography}%

\end{document}